# Size Dependent Ternary Halide Solid Solutions in Perovskite Nanocrystals

Shai Levy[1†], Lotte Kortstee[2†], Georgy Dosovitskiy[1], Emma H. Massasa[1], Yaron Kauffmann[1], Juan Maria García-Lastra[2], Ivano E. Castelli[2], and Yehonadav Bekenstein[1]
1 Department of Materials Science and Engineering and The Solid-State Institute, Technion-Israel Institute of Technology, 32000 Haifa, Israel
2 Department of Energy Conversion and Storage (DTU Energy), Technical University of Denmark, Agnes Nielsens Vej 301, DK-2800 Kongens Lyngby, Denmark
[†]These authors contributed equally to this work
Author Address: bekenstein@technion.ac.il

**Abstract:** Crystalline solid solutions incorporate guest atoms by substituting host lattice sites up to a solubility limit dictated by solute–host similarity. Solid solutions enable tuning various material properties, such as the optoelectronic behavior of halide perovskites. In bulk, incompatibility of Cl:I halide mixtures restricts exploration throughout the ternary halide Cl:Br:I compositional range. However, solubility is extended in nanocrystals which better accommodate a wider range of ions within their lattice. Through high-throughput synthesis and spectroscopic characterization of over 3000 samples, along with density functional theory and cluster expansion models, we determine the solubility boundaries of ternary halide perovskite nanocrystals and demonstrate their extended size-dependent miscibility. Smaller nanocrystals, with sufficient Br content, stabilize the Cl:Br:I solid solutions, suppress planar stacking fault defects and prevent halide segregation.

## 1. Introduction

Quantum dots of halide perovskites (ABX$_3$, X=Cl,Br,I) are a unique platform for quantum optical effects, ranging from single photon emission in individual nanocrystals (NCs) [1–3] to correlative multiphoton superfluorescence in nanocrystal ensembles[4,5]. To effectively utilize these quantum effects, it is possible to tune their optical band gap and quantum confinement by changing their halide composition [1]. Uniform anion distribution in these multicomponent crystals requires the formation of a substitutional solid solution, in which the occupancy of each lattice site is purely statistical. Formation of binary halide perovskites with compositions such as Cl:Br or Br:I is therefore a common strategy for tuning their band gap throughout the visible spectrum, either during their synthesis[6] or in a post synthetic anion exchange process[7–9]. This tunability makes both bulk crystals and nanocrystals of halide perovskites promising materials for various optoelectronic applications such as light emitters[6,10,11], highly efficient photovoltaics[12,13] and ultrafast detectors[14]. Therefore, the next logical step in modifying the optoelectronic properties of perovskites is to utilize more complex compositions which combine all three halides (Cl:Br:I) in the same structure. Ternary halide perovskites were shown to improve the performance of bulk perovskite solar cells, preventing the undesired phenomenon of halide segregation[15–18]. Despite such advantages, many compositions of ternary halides are thermodynamically unstable for bulk crystals[19,20]. Halide mixtures which include both Cl and I suffer from an inherent poor stability[21], related to the fundamental requirements of crystalline solid solutions.

Stability of any ionic crystal is dependent on ionic sizes, as was shown by Pauling in the 1920s[22]. Anion to cation size ratio determines the coordination number for each ion, enabling the most efficient overall packing of the crystal. In perovskite materials, criteria related to Pauling rules such as Goldschmidt tolerance factor[23] and octahedral factor[24] are often considered as local rules in determining the stability of the crystal. However, these rules are insufficient for predicting the stability of crystals with a mixture of ions substituting the same lattice site. The requirements for substitutional solid solutions were formulated by Hume-Rothery in 1926 [25], empirically observed in metals and later in ionic crystals, both dielectrics and semiconductors [26]. To satisfy the conditions for a substitutional solid solution, the ions should have similar size ($\Delta r<15\%$, for r the ionic radius), electronegativity, and preferred crystalline structure. Using these guidelines with the appropriate ionic sizes for cesium lead halide perovskites[27], Cl:Br ($\Delta r=8\%$) or Br:I ($\Delta r=12\%$) binary halide compositions should show full miscibility, while Cl:I ($\Delta r=22\%$) mixtures are incompatible, and have a low solubility limit of below 5at%[21,28].

Solubility in crystalline solid solutions is enhanced at the nanoscale, as the high surface to volume ratio helps to accommodate a wider range of ions within the lattice. This size-dependent stabilization, was observed to transform immiscible bulk metallic systems into fully miscible alloys in nanocrystals[29–31]. Additionally, a recent report by Manna et.al. showed that perovskite nanocrystals are able to accommodate ethylammonium as an A-site cation, breaking the expected limitation predicted by Goldschmidt tolerance factor [32]. This observation opens the possibility to stabilize otherwise thermodynamically unfavorable halide mixtures in perovskite nanocrystals.

## 2. Results and Discussion

We began by comparing the optical properties of different binary halide perovskite NC systems undergoing an anion exchange process. We used $CsPbBr_3$ NCs and measured their photoluminescence (PL) and absorption spectra after anion exchange with different loadings of either $PbCl_2$ or $PbI_2$ 5mM solutions, as shown in Fig.1a-b. In the presence of a guest halide, diffusion and anion substitutions occur, resulting in NCs of binary halide composition with shifted PL peak and absorption onset. In both reactions, the substitution of Br to either Cl or I creates a solid solution, showing a gradual PL shift, which is linearly dependent on the X-site composition. The PL peak position is therefore a good indicator for the effective halide composition of the perovskite [7].

Initiating the anion exchange process from $CsPbCl_3$ NCs instead of $CsPbBr_3$ NCs through adding varying amounts of $PbBr_2$ or $PbI_2$, showed a different behavior. In Cl to Br exchange shown in Fig.1c, the PL peak gradualy red shifts with the increasing Br concentration, indicating a successful anion exchange and the formation of a solid solution. However, addition of I to the $CsPbCl_3$ NCs shown in Figure 1d, shows significant quenching in violet $CsPbCl_3$ emission with slight red shift for low I loadings. With higher amounts of $PbI_2$, a new red emission peak emerges, which further shifts with the addition of I from 625nm to 675nm. This behavior is the result of the low miscibility in the Cl:I binary system due to their incompatability. It was shown previously that up to 5at% of I can be incorporated especially in the surfaces of the $CsPbCl_3$ NCs leading to slight PL shift. Above this level, a new iodide dominated composition is formed alongside the $CsPbCl_3$ NCs as indicated by the red light emission[21].

As binary Cl:I mixtures showed limited miscibility, we set to examine the behavior of ternary halide perovskite NCs, to explore the effects of intermediate-sized Br ions on the Cl:I system. By anion exchange of CsPbCl$_3$ NCs with various loadings of both PbBr$_2$ and PbI$_2$, with a total exchange solution volume of 45µl, the abrupt shift from 480nm Cl:Br dominated composition to 650nm I dominated composition is still observed (Fig.1e), indicating the miscibility gap. However, above a certain level of Cl exchange, as shown for total loading of 120µl in Fig.1f, the PL shows a gradual shift in the region matching with Br:I rich compositions, indicating full miscibility behavior with sufficient Br content. The mappings of the PL peak energy and PL quantum yield (PLQY) presented in Fig.1g-h, help to determine the required amount of Br to bridge the Cl:I miscibility gap. Above PbBr$_2$ loading of 50µl, corresponding to 40at%, the PL peak shifts gradually from blue to red with a higher I loading, and the overall PLQY is higher as an indicator of a reduced number of defects in the NCs after this exchange process.

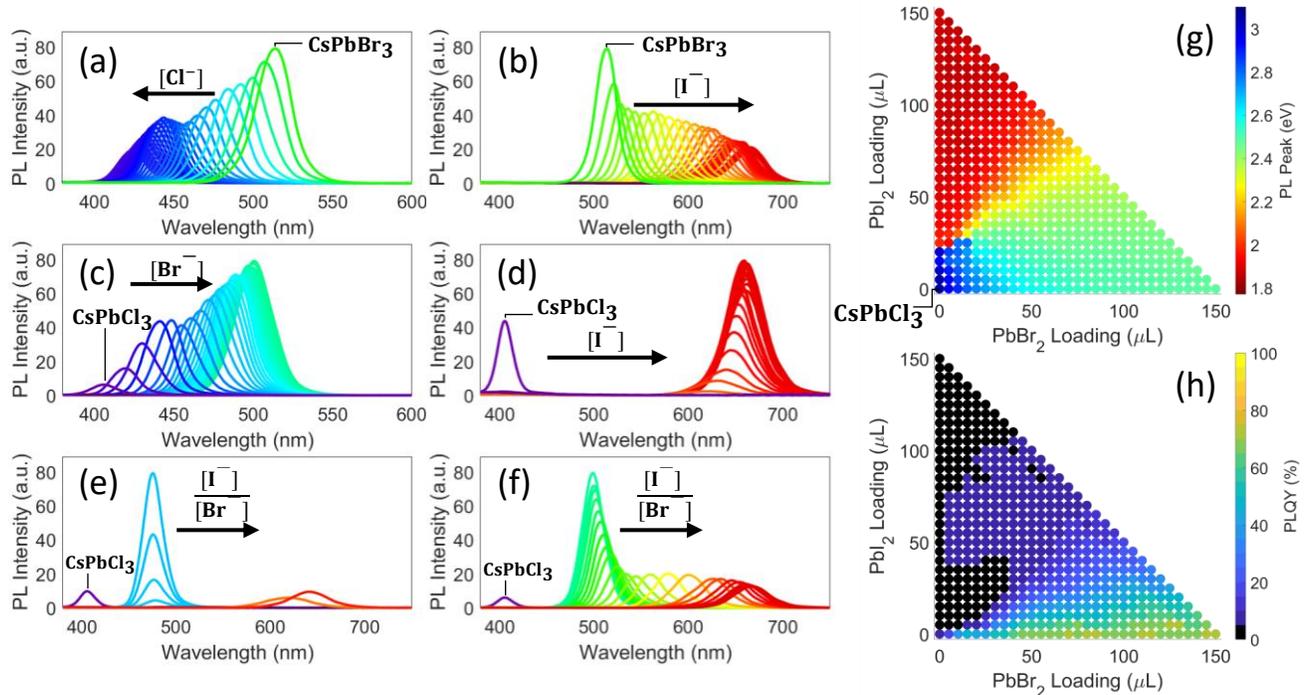

**Figure 1. Photoluminescence of halide exchanged nanocrystals. (a-b)** PL spectra of binary halide perovskite NCs produced from CsPbBr$_3$ NCs after anion exchange process with various loadings of either (a)PbCl$_2$ or (b)PbI$_2$. Both exchanges result in fully miscible solid solution systems with a continuous shift of the PL according to the halide composition. **(c-d)** PL spectra of binary halide perovskite NCs produced from CsPbCl$_3$ NCs after anion exchange process with various loadings of either (c)PbBr$_2$ or (d)PbI$_2$. Exchange with Br presents full miscibility, while the exchange with I shows a significant miscibility gap. **(e-f)** PL spectra of ternary halide perovskite NCs produced from CsPbCl$_3$ NCs after anion exchange with varying the loading of both PbBr$_2$ and PbI$_2$, with total exchange solution loading of (e)45µl or (f)120µl. **(g-h)** Emission maps of CsPbCl$_3$ NCs after anion exchange process with varying loading ratio of PbBr$_2$ and PbI$_2$ solutions. (g) presents the PL peak energy and (h) the relative PLQY, using quinine sulfate as a standard.

To computationally investigate the thermodynamic stability throughout the ternary halide compositional range, we used a Cluster Expansion (CE) model[33]. The CE method relies on mapping material properties by establishing a relationship between a physical quantity, in our case E the total energy of a structure as obtained by Density Functional Theory (DFT), and a configuration **σ**, which is an N dimensional vector containing N atomic site variables. By dividing the crystalline unit cell into atomic clusters of various sizes (with up to 4-body interactions), we establish a correlation function $\phi_\alpha$ of a cluster with symmetry α, which encapsulates the characteristics of a cluster through the associated orthogonal single-site basic functions it is made of. This correlation function describes the lattice in terms of occupation of a specific atomic species per site. Through the correlation function, the physical quantity E(**σ**) can be expressed as follows[34]:

$$(1) \; E(\pmb{\sigma}) = \sum_{\alpha} \widetilde{J}_{\alpha} \phi_{\alpha}$$

Where $\widetilde{J}_{\alpha}$ is the effective cluster interaction (ECI) per atom associated with the cluster α. In essence, this means that the ECI values hold the energetic contribution, or weight, of a certain cluster to the total energy of the structure. Through methods described in the Supporting Information (S.I), we built an autonomous CE model using various structural interactions[35] and found the ground state of 462 bulk structures spanning the ternary halide compositional range. The formation energy of each composition, referenced against the appropriate single halide phases, is presented in Figure 2a. This CE model predicts a large area of stability for ternary halide perovskites, showing an energy gain of up to ~5-10 meV/atom. Moreover, the CE model also predicts a region of instability for compositions rich with Cl:I (for X-site composition of <20at%Br). This prediction shows the inherent thermodynamic instability in certain halide solid solutions, arising from their incompatibility.

To investigate if this predicted bulk thermodynamic instability holds for nanocrystals, we performed a high-throughput anion exchange on 12.4nm $CsPbBr_3$ NCs with various loadings of $PbCl_2$ and $PbI_2$. In this case, we can use optical calibration, based on the binary Cl:Br and Br:I compositions, to convert the loading of the anions in the medium into an estimated X-site composition. This calibration, described in the S.I, is based on the linear change of the optical band gap energy with the composition in the binary halide mixtures[7]. First, we considered the optical band gap of each sample relative to their anion loading to find the effective substitution percentage of the halide sites in the NCs. Then, we extrapolate these curves to the ternary halide case. This method of composition estimation is additionally supported by scanning electron microscopy energy dispersive spectroscopy (SEM-EDS) elemental analysis measurements on selected samples as described in the S.I.

Measurements of the photoluminescence quantum yield (PLQY) throughout the ternary halide composition range are shown in Fig.2b. As PLQY is the ratio between the emitted and absorbed photons by the samples, it can be used as an indicator for crystalline defects which can act as traps leading to non-radiative relaxation[36]. As we expected from the CE model, we observe an emission quenching in a wide range of compositions that are rich with Cl:I. Additional evidence for the formation of defects in these ternary halide NCs is decreased radiative lifetime, measured by Time-resolved PL for Cl:I NCs with low Br content (Figure S4). These results suggest that incompatible halide mixtures have higher amounts of defects which reduce their PLQY significantly, even in compositions not expected to be unstable by the CE model.

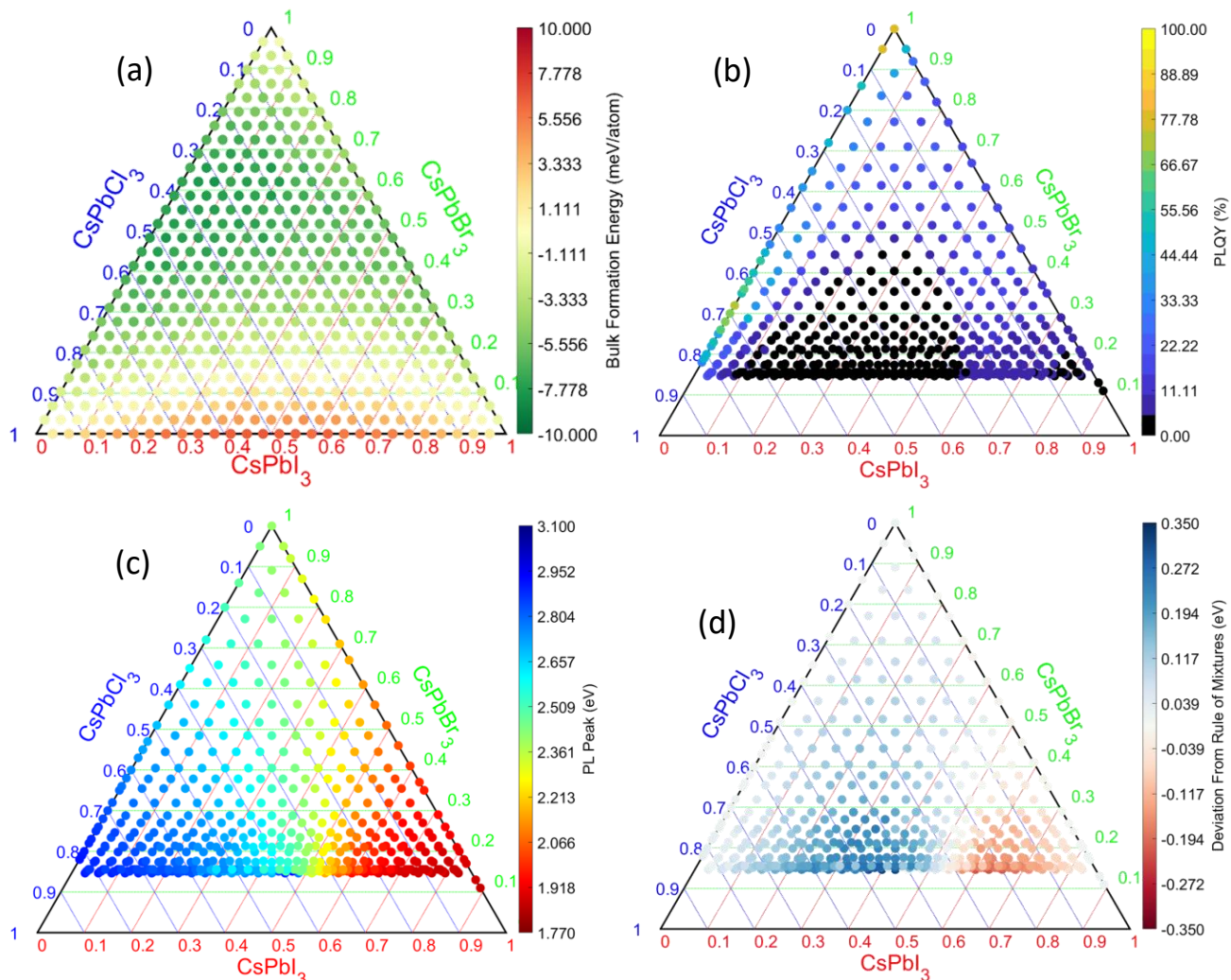

**Figure 2. Stability and optical properties of ternary halide perovskite nanocrystals. (a)** Cluster expansion DFT calculations of bulk formation energy throughout the ternary halide composition range, showing inherent instability of Cl:I rich halide mixtures. **(b)** Relative PLQY using Quinine sulfate dye, showing increased defect formation reducing PLQY in Cl:I rich halide mixtures. **(c)** Emission maximum of ternary halide perovskite nanocrystals. **(d)** Deviation from ideal solid-solution rule of mixtures in the PL peak, presented in panel (c).

Additionally, unlike the case of anion exchange produced from $CsPbCl_3$ NCs, the ternary halide NCs obtained from $CsPbBr_3$ show a gradual PL shift that spans over almost the entire expected range from pure $CsPbI_3$ to $CsPbCl_3$ (Fig.2c). To examine if these ternary halide perovskites manage to form a uniform solid solution, we compared the PL peak of each composition to the PL peak expected by the rule of mixtures which is a weighted average of the single halide phases. Deviations between our measured results and the expected PL peak values from the rule of mixtures are presented in Fig.3d. We observed that ternary halide NCs with compositions either close to binary halide compositions, or with Br concentration above 40at%, show small deviation from the rule of mixtures (below 40meV), as expected from uniform solid solutions. Meanwhile, ternary halide compositions with lower amounts of Br, do show significant deviation from uniform solid solution mixing in two domains. Systems with high amount of I are strongly red-shifted, and systems with higher Cl are blue-shifted relative to the expected value. This is a signal of nonuniform halide mixing due to a breaking in the solid solution through halide segregation, despite the measured composition change of the NCs. As a result, one of the halides is less represented in the exciton medium and energy levels of the semiconductor. Therefore, we can use the deviation from a rule of mixture to determine the cases of formation or breaking of the ternary halide solid solution alloys.

The ability of the crystals to accommodate the different halides within a uniform solid solution is expected to be related to their size. By tuning the synthesis conditions of CsPbBr$_3$ NCs, we obtained monodispersed NCs with various average sizes of 12.4nm, 7.8nm, 6.3nm, and 4.7nm, as shown by transmission electron microscopy (TEM) in Fig.3a-d. Each sample underwent high-throughput anion exchange process, followed by spectroscopic characterization. Their PLQY presented Fig.3e-h are used as a marker of defect formation from the halide mixing. Larger NCs showed a wider range of halide loadings with emission quenching, while smaller NCs had a significantly reduced emission quenching regions. The extended compositional range maintaining high PLQY in smaller NCs demonstrate the ability of the NC surface to stabilize otherwise unfavorable ternary halide solid solutions.

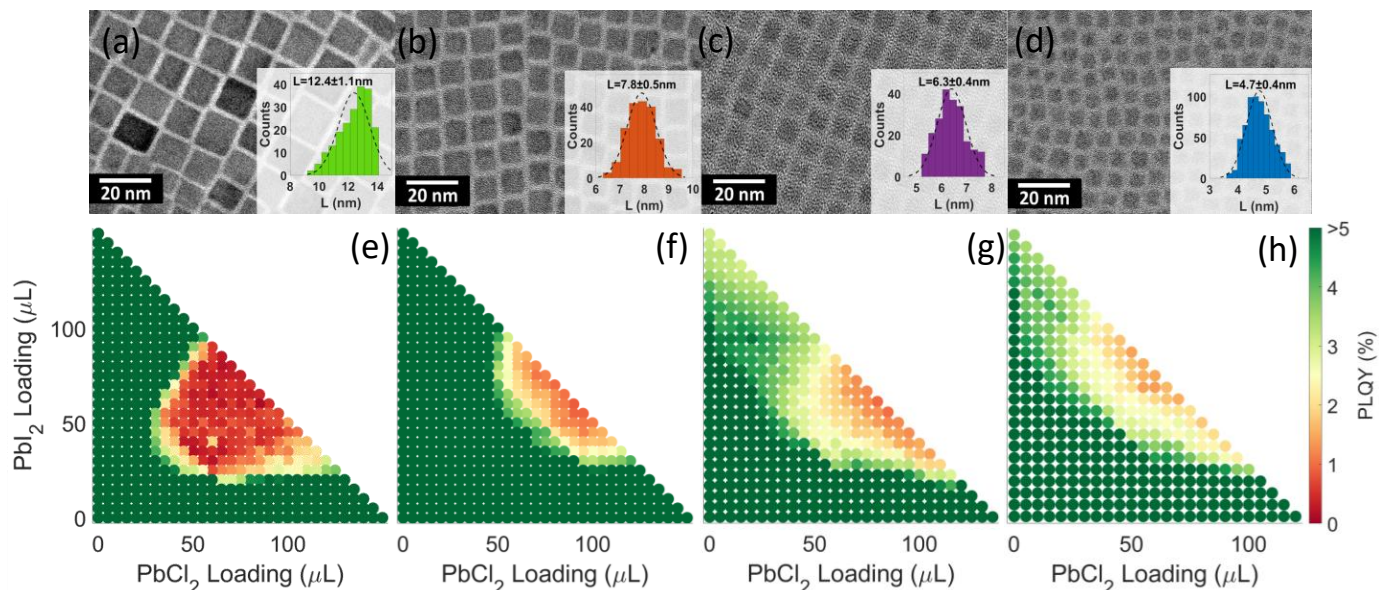

**Figure 3. Nanocrystal size effects on the ternary halide solid solutions. (a-d)** TEM micrographs and size distribution histograms of native CsPbBr$_3$ perovskite nanocrystals with average sizes of (a) 12.4nm, (b) 7.8nm, (c)6.3nm, and (d) 4.7nm. **(e-h)** PLQY mapping of ternary halide perovskite nanocrystals, with different PbCl$_2$ and PbI$_2$ loadings, by anion exchange using native CsPbBr$_3$ with sizes matching (a-d) respectively.

To further explain how the surface influences the thermodynamic stability of ternary halide perovskites, we performed DFT calculations on a series of surface models. These models aim to explain how Br promotes the incorporation of guest halides into a host lattice of a given halide composition. For several selected ternary halide compositions (described in Fig. 4a), we constructed a set of surface structures. These structures differ only by the number of atomic layers from the surface to which the guest halide had penetrated and incorporated into, as illustrated in Fig. 4b (and Fig. S11). In a stepwise manner, we introduced the guest halide into deeper layers of the crystal, with bottom layers designated as fixed layers which remained with pristine host lattice composition. All surfaces were modeled along the {100} orientation with a PbX$_2$ termination, consistent with the surface observed in these NCs[37].

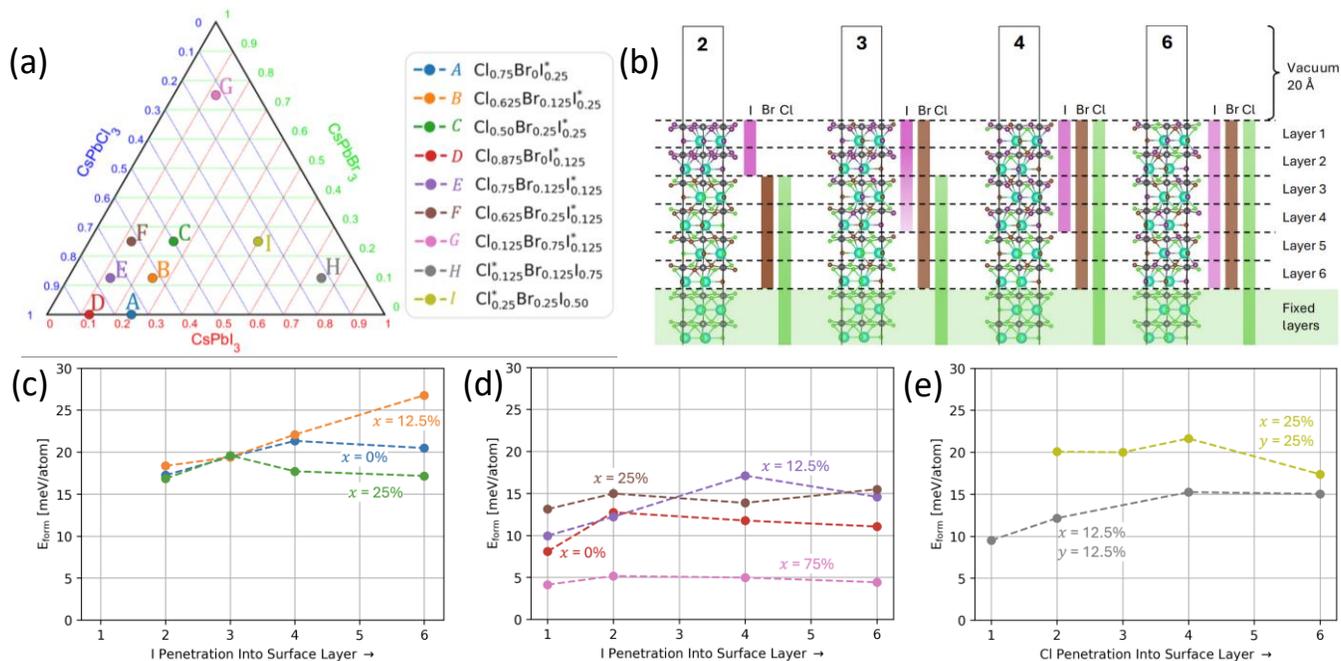

**Figure 4. DFT surface models of ternary halide perovskites (a)** Compositional map of selected ternary halide compositions used in the DFT surface models with the incorporated guest halide is marked by *. **(b)** Scheme of the surface model simulation used for composition C. The numbers 2, 3, 4 and 6 indicate atomic layer from the surface to which the guest halides are incorporated. The colored bars to the side of the slab signify the presence of each halide species in specific crystalline layer. **(c-e)** Surface formation energy for different incorporation of guest halides into the crystalline layers of the host lattice, matching in color to A-I compositions. (c) Cases of 25% of I stabilized by $x$ Br in the host lattice, (d) cases of 12.5% I stabilized by $x$ Br in the host lattice, and (e) cases of $x$ Cl stabilized by y Br in the lattice.

The formation energies for these mixed compositions were referenced against surface energies of single halide compositions as shown in Figure 4c-e. Two key trends emerge from these surface models, we observe that in the absence of Br as a stabilizing agent, lower energetics are obtained for structures with the guest halides segregated to the surface layer of the crystal (Figure 4b in blue and Figure 4c in red). Additionally, Br content is stabilizing the full incorporation of the guest ions in the crystal. This can be observed by comparing compositions with Br content of 12.5at% and 25at%, where the configuration of the latter shows the lowest energy when guest ions are incorporated into the deeper layers of the crystals (Figure 4b in orange and green and Figure 4c in purple and brown). Similar trends are observed by models with I-rich poor host lattice composition. Moreover, if the host lattice consists mostly of Br, full incorporation of the guest species in the lattice is predicted (Figure 4c in pink). These results offer us a thermodynamic explanation for the breaking of a uniform solid solution and halide segregation in cases without sufficient Br acting as intermediate stabilizing ions. By simulating energy of surfaces and other lattice sites which are unsaturated chemically, we better understand the experimental size dependency of the ternary halide solid solutions.

In order to characterize the structure of the ternary halide perovskite NCs we performed X-ray diffraction (XRD) measurements on several ternary halide compositions, as shown in Fig.5a. All samples showed a diffraction pattern matching to the perovskite structure, as confirmed by ICDD references 01-084-0437 for $CsPbCl_3$, 00-054-0752 for $CsPbBr_3$, and 04-016-2301 for $CsPbI_3$. Close inspection of the {200} reflection peaks shows a single peak for each sample in the predicted range of diffraction angles. This confirms that all samples are not composed of several significantly different halide compositions which should have different lattice parameters. Additionally, we observe that the measured lattice parameters, as the PL peak, do not follow the rule of mixtures or Vegard's law in accordance with their halide composition. This is an additional sign that some halide compositions break from a uniform solid solution in some respects.

To visualize both uniform ternary halide solid solution and cases of segregated halides, we conducted quantitative contrast analysis of high angle annular dark field (HAADF) scanning transmission electron microscopy (STEM) micrographs. Since these NCs are highly sensitive to e-beam irradiation EDX mapping is not possible. To overcome this limitation we used the StatSTEM software[38] to analyze the contrast of the PbX atomic columns in the micrographs and calculated their electron scattering cross-section. As heavier elements have higher scattering cross-sections, we can attribute PbX atomic columns with higher intensity to more I and Br than Cl. In smaller NCs, such as the 6.6nm NC presented in Fig.5b-c, the lattice seems pristine. On the other hand, some larger NCs, such as the 10.3nm NC presented in Fig.5d-e, have distinct planar defects. These defects are stacking faults or missing planes in the crystal structure and were previously reported for perovskite NCs containing both Cl:I[39]. In some cases, as presented in the TEM micrograph at the inset of Fig.5d, multiple staking faults in one NC arrange to form a frame, separating core region to an outer shell layer. As we observe these staking faults after the process of anion exchange it is reasonable that their formation is a pathway for relaxation of unstable ternary halide compositions. Using this mapping for the halide distributions of halides, we observe uniform halide mixing in the smaller NC, and heavier halide segregation to the stacking fault defect with a broken solid solution in the larger NC. This highlights the importance of the NC size in stabilizing the ternary halide solid solutions, preventing undesired halide segregation and defects which are detrimental for their optical properties.

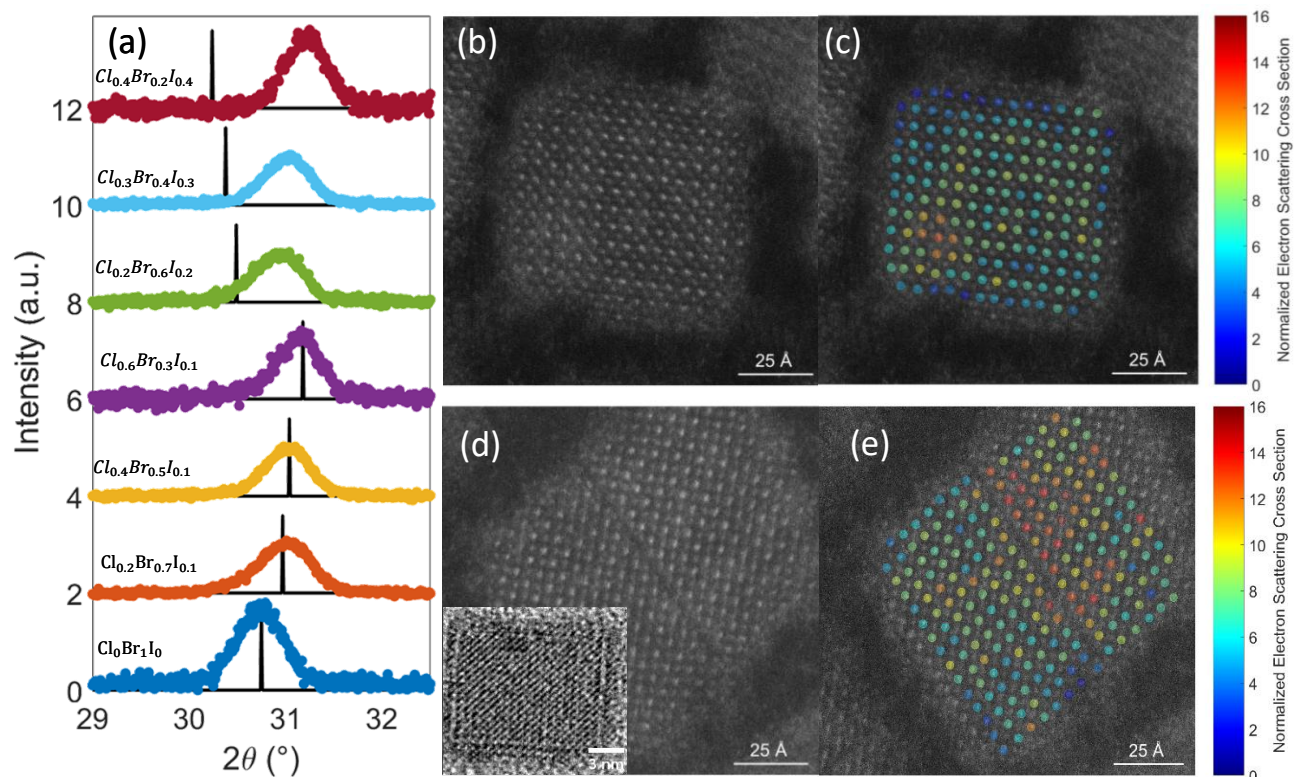

**Figure 5. Structural characterization of ternary halide perovskite nanocrystals. (a)** XRD diffraction {200} reflections of ternary halide perovskite nanocrystals with different halide compositions, with black lines representing the expected {200} reflection for these compositions by Vegard law based on ICDD references. **(b)** HAADF STEM micrograph of a uniform 6.6nm ternary halide perovskite nanocrystal. **(c)** StatSTEM Z-contrast analysis of (b), showing uniform distribution of halides in the lattice. **(d)** HAADF STEM micrograph of 10.3nm ternary halide perovskite nanocrystal with a stacking fault defect. Inset: TEM micrograph of halide perovskite nanocrystal with multiple stacking fault defects forming a core-shell configuration. **(e)** StatSTEM Z-contrast analysis of (d), showing segregated halides near the stacking fault defect and nanodomains with different relative composition.

## 3. Conclusion

We investigated the boundaries of halide miscibility and solid solution formation in perovskite nanocrystals using a high-throughput anion exchange and spectroscopy. Ionic compatibility and nanocrystal size govern the uniformity of ternary halide mixtures. Compositions with Br content above 40at% at the X-site form uniform solid solutions, whereas Cl:I rich compositions show photoluminescence quenching, reduced radiative lifetimes, and deviations from the rule of mixtures, indicating halide segregation and defect formation, as predicted by our CE model. Smaller nanocrystals better accommodate incompatible halide mixtures, displaying fewer defects and higher photoluminescence quantum yields. This is supported by DFT surface calculations, showing preference of incompatible Cl and I guest ions to reside at the surface layers of the crystals, unless 12.5at% to 25at% of Br, which is sufficient to stabilize the Cl:I incorporation to the crystal. Additional structural analysis confirmed that halide segregation and stacking fault defects occur primarily in larger NCs, emphasizing the role of nanocrystal size in stabilizing the ternary halide compositions. Using these results, we envision using homogeneous ternary halide compositions in tuning the quantum optical properties of perovskite nanocrystals.

**Acknowledgements**

We thank Noa Goeli and Noa Zilberman for their help in this research as part of their undergraduate project. We thank Jessica Nachamie for her help in data analysis. **Funding:** This project is supported by the European Union's Horizon 2020 research and innovation program under grant agreement No 949682- ERC. We thankfully acknowledge RES resources provided by Barcelona Supercomputer Center in MareNostrum5 to project QHS-2025-1-0036. Additionally, we acknowledge support from the Novo Nordisk Foundation Data Science Research Infrastructure 2022 Grant under No. NNF22OC0078009: "A high-performance computing infrastructure for data-driven research on sustainable energy materials". L.K. and I.E.C. acknowledge support from DTU through the Alliance Ph.D. Research Project "AI-accelerated Discovery of Self-healing Lead-free Metal-halide Perovskites for Solar Energy Conversion". G.D. acknowledges the support from the Council for Higher Education and the Center for Integration in Science of the Ministry of Aliyah and Integration of Israel. **Competing interests:** The authors declare no competing interests. **Data and materials availability:** All data is available in the manuscript or the Supporting Information.

## Author Information

### Corresponding Author

**Yehonadav Bekenstein** − Department of Materials Science and Engineering and the Solid-State Institute, Technion − Israel Institute of Technology, 32000 Haifa, Israel; orcid.org/ 0000-0001-6230-5182; Email: bekenstein@technion.ac.il

### Authors

**Shai Levy** - Department of Materials Science and Engineering and the Solid-State Institute, Technion − Israel Institute of Technology, 32000 Haifa, Israel; orcid.org/0000-0001-6376-0486
**Georgy Dosovitskiy** - Department of Materials Science and Engineering and the Solid-State Institute, Technion − Israel Institute of Technology, 32000 Haifa, Israel.
**Lotte Kortstee** - Department of Energy Conversion and Storage (DTU Energy), Technical University of Denmark, Agnes Nielsens Vej 301, DK-2800 Kongens Lyngby, Denmark; https://orcid.org/0009-0003-0822-5706
**Emma H. Massasa** - Department of Materials Science and Engineering and the Solid-State Institute, Technion − Israel Institute of Technology, 32000 Haifa, Israel; https://orcid.org/0000-0002-6803-5379
**Yaron Kauffmann**- Department of Materials Science and Engineering and the Solid-State Institute, Technion − Israel Institute of Technology, 32000 Haifa, Israel.
**Juan Maria García-Lastra** - Department of Energy Conversion and Storage (DTU Energy), Technical University of Denmark, Agnes Nielsens Vej 301, DK-2800 Kongens Lyngby, Denmark; https://orcid.org/0000-0001-5311-3656
**Ivano E. Castelli** - Department of Energy Conversion and Storage (DTU Energy), Technical University of Denmark, Agnes Nielsens Vej 301, DK-2800 Kongens Lyngby, Denmark; https://orcid.org/0000-0001-5880-5045

### Author Contributions

S.L. and Y.B. designed the experiments. S.L. performed robotic synthesis and spectroscopic measurements. G.D. managed the robotic experimental setup. S.L. and E.M performed the structural characterization. L.K., J.M.G.L and I.E.C. designed and performed the DFT calculations and trained CE model. S.L. analyzed the results and wrote the manuscript with help from the other co-authors.

*Supporting Information for*

# Size Dependent Ternary Halide Solid Solutions in Perovskite Nanocrystals

Shai Levy[1†], Lotte Kortstee[2†], Georgy Dosovitskiy[1], Emma H. Massasa[1], Yaron Kauffmann[1], Juan Maria García-Lastra[2], Ivano E. Castelli[2], and Yehonadav Bekenstein[1]

1 Department of Materials Science and Engineering and The Solid-State Institute, Technion-Israel Institute of Technology, 32000 Haifa, Israel
2 Department of Energy Conversion and Storage (DTU Energy), Technical University of Denmark, Agnes Nielsens Vej 301, DK-2800 Kongens Lyngby, Denmark
[†]These authors contributed equally to this work
Author Address: bekenstein@technion.ac.il
Corresponding author: Yehonadav Bekenstein, bekenstein@technion.ac.il

**The file includes:**
Materials and Methods
Supplementary Figures. S1 to S13
References

## Materials and Experimental Methods

**Chemicals:**

Cesium carbonate (99.9%, Aldrich), decane (99%, Aldrich), hexane (A.R., Aldrich), lead bromide (99.998%, Aesar), lead chloride (99.999%, Aldrich), lead iodide (99%, Aldrich), octadecene (90%, Acros), oleic acid (90%, Aldrich), oleylamine (98%, Aldrich), quinine sulfate (98%, Angene), sulfuric acid (95%, Aldrich), and zinc bromide (99.9%, Aesar).

All chemicals were used as purchased without further purification.

**Synthesis of $CsPbX_3$ Nanocrystals:**

First, Cs-oleate precursor was prepared in a 25 mL three-necked round-bottomed flask by dissolving $Cs_2CO_3$ (0.25 g) in a mixture of oleic acid (0.8 g) and octadecene (7 g) at 150 °C for 10 min under $N_2$ atmosphere in a Schlenk line. The precursor solution of $PbX_2$ was prepared by dissolving $PbBr_2$ (75 mg) in a mixture of octadecene (5 mL), oleic acid (2 mL), and oleylamine (2 mL) in a 25 mL three-necked round-bottomed flask under $N_2$ atmosphere at 120 °C for 10 min. After the temperature of the precursor solution was set to 200 ºC, 0.4 mL of Cs precursor solution was injected to initiate the reaction. The reaction was quenched after a few seconds in an ice-water bath. The product was centrifuged twice at 9000 rpm, redispersed in clean hexane or decane, and again at 3500 rpm. Synthesis of $CsPbBr_3$ NCs with tuneable sizes was conducted by addition of varying amount of $ZnBr_2$ (0-414mg) to the $PbX_2$ solutions and by changing the injection temperature as reported by Dong et al.(*1*).

**High-throughput halide exchange:**

High-throughput halide exchange experiments were performed using Opentrons OT-2 liquid handling robot, equipped with single-channel 300 µl and 20 µl pipettes (Opentrons P300 and P20 GEN2) and capable of dispensing liquids with 1-5 µl precision. Lead halide stock solutions in decane were prepared by dissolving $PbX_2$ (139.0 mg $PbCl_2$, or 183.5 mg $PbBr_2$, or 230.5 mg $PbI_2$), in 16 ml decane, 2 ml oleic acid and 2 ml oleylamine. These stock solutions were diluted 1:100 in decane to obtain 0.25mM concentration of the $PbX_2$. The stock solutions were dispensed into 96-well plates (Corning Costar, flat clear bottom, black wall polystyrene plate, with total well volume of 360µl) and set on a heater-shaker (Opentrons Heater-Shaker Module GEN 1). Each well was loaded with the desired loading of 0.25mM $PbX_2$ stock solutions in the range of 5-150 µl, topped to 150 µl total volume with decane, and mixed. Then, 50 µl of NCs solutions ($CsPbCl_3$, or $CsPbBr_3$, or $CsPbI_3$) were added to each well, keeping a consistent total volume of 200 µl, as the last step in the process to minimize differences in NCs dwelling time in the ion exchange medium. Then the plate was mixed for 1 minute, left for 5 more minutes, and mixed again, and was sent for high-throughput spectroscopic measurements.

**High-throughput UV–vis absorption, and PL measurements:**

Steady-state absorption and emission spectra of the samples in the 96-well microplates were measured using an Agilent BioTek Synergy H1 hybrid multimode reader, with a xenon lamp (Xe900) source. 200 µl of clean decane dispensed to well H12 was used as a blank sample for the NC samples.

PLQY was measured relatively by comparison to 200 µl of quinine sulfate standard fluorescent dye dissolved in 0.05M $H_2SO_4$ aqueous solution (dispensed to well G12, with distilled water as dedicated blank in well F12). PLQY was calculated using the method described by Brouwer (*2*).

**Transmission Electron Microscopy (TEM):**

One drop of dilute nanocrystal solution in hexane was cast onto a TEM grid (carbon film only on a 300mesh copper grid). The samples were observed in TEM mode with a Thermo Fisher/FEI Tecnai $G^2$ T20 S-Twin $LaB_6$ TEM operated at 200 K with a Gatan Rio9 CMOS camera.

High-resolution imaging, diffraction patterns acquisition, and elemental mapping were done in a Thermo Fisher/FEI Titan-Themis double-Cs-corrected HR-STEM operated at 300 kV and equipped with Ceta2 4K × 4K camera (for TEM mode) and Bruker Dual-X EDX detectors for STEM-EDX chemical mapping. The high-resolution STEM micrographs were acquired using a high-angle annular dark-field (HAADF) STEM detector with a collection angle range of 93–200° mrad and beam convergence of 21° mrad.

## X-ray diffraction (XRD):

The nanocrystal solution in hexane was dropped-casted onto a glass substrate (rectangular microslides, 76 × 26 [mm]), and the X-ray beam was focused on the resulting film. Measurements were taken using a Rigaku Smart-Lab 9 kW high-resolution X-ray diffractometer equipped with a rotating anode X-ray source. We used the "Glancing mode" (grazing angle) method (2-theta), which is suitable for measuring thin films, with a 1.54 Å (Cu Kα) wavelength. The X-ray source was fixed on ω = 0.4°, and the detector was moved in the range of 2θ = 20–90°.

## Additional Experimental Results

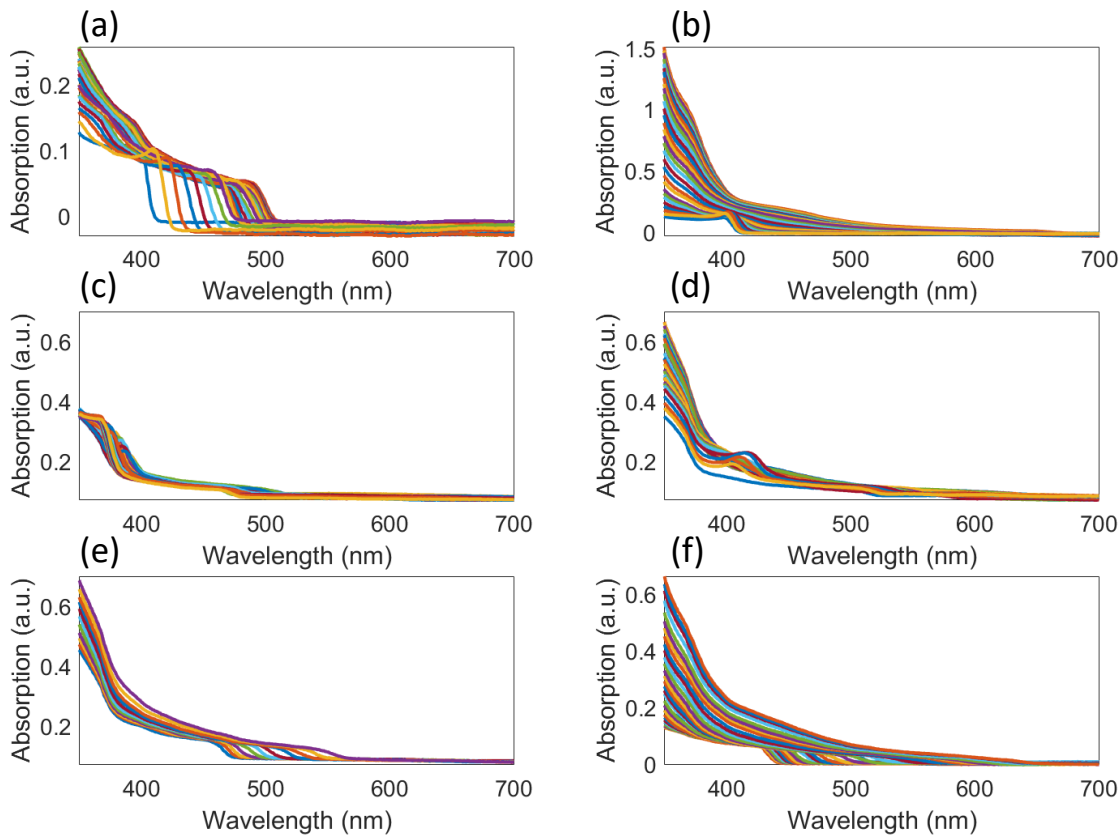

**Figure S1. (a-b).** Optical absorption spectra of binary halide perovskite NCs using $CsPbCl_3$ NCs after anion exchange process with various loadings of either (a)$PbBr_2$ or (b)$PbI_2$. **(c-d).** Optical absorption spectra of binary halide perovskite NCs using $CsPbBr_3$ NCs after anion exchange process with various loadings of either (c)$PbCl_2$ or (d)$PbI_2$. **(e-f).** Optical absorption spectra of ternary halide perovskite NCs using $CsPbBr_3$ NCs after anion exchange process with various loadings of both $PbBr_2$ and $PbI_2$ with total loading volume of (e) 140μl (f) 60μl.

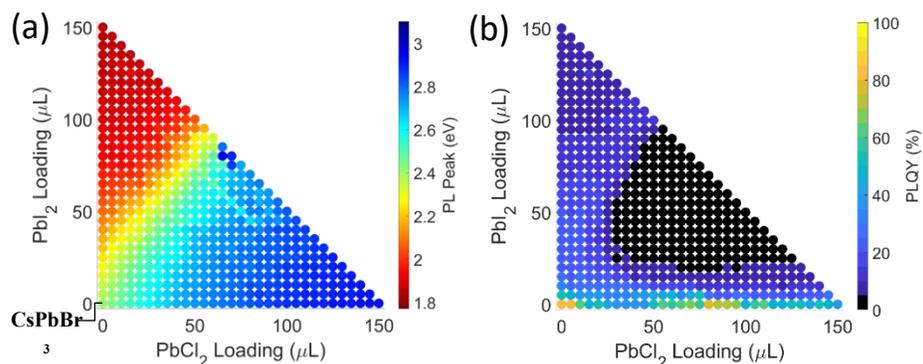

**Figure S2. (a-b)** Emission maps of 12nm CsPbBr$_3$ NCs after anion exchange process with varying loading ratio of PbCl$_2$ and PbI$_2$ solutions. (a) present the PL peak energy and (b) the relative PLQY, using quinine sulfate as a standard.

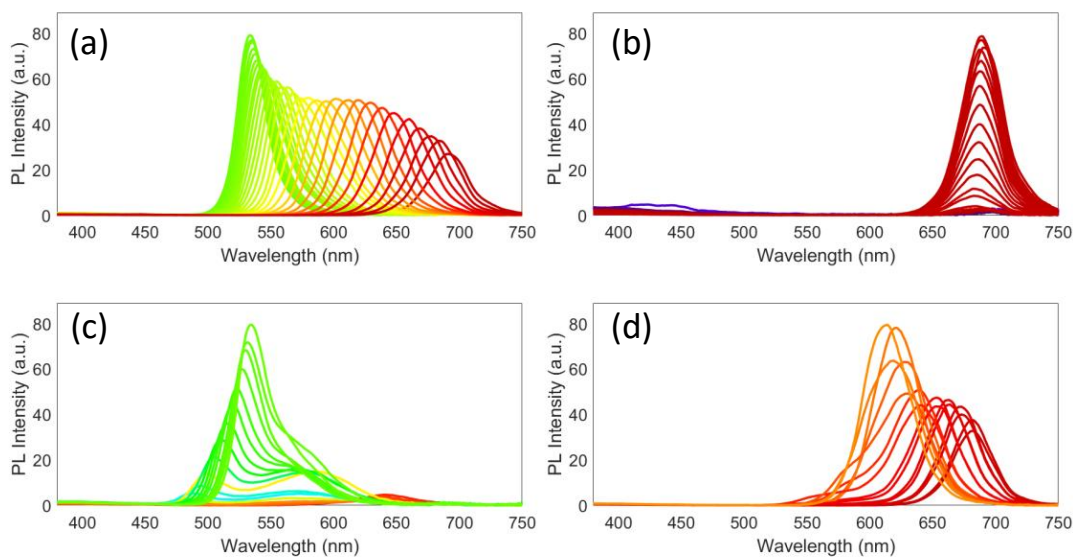

**Figure S3. (a-b)** PL spectra of binary halide perovskite NCs using CsPbI$_3$ NCs after anion exchange process with various loadings of either (a)PbBr$_2$ or (b)PbCl$_2$. Exchange with Br presents full miscibility, while an exchange with Cl shows a significant miscibility gap. **(c-d)** PL spectra of ternary halide perovskite NCs using CsPbI$_3$ NCs after anion exchange with varying the loading of both PbBr$_2$ and PbCl$_2$, with total loading of exchange solutions of (c)130μl or (d)45μl.

**Time resolved photoluminescence (TRPL):**

Radiative lifetime measurements were performed on 96-well microplates using Edinburgh instruments FLS1000 spectrometer coupled to Nikon Eclipse Ni-U optical microscope in time-correlated single-photon counting (TCSPC) mode under pulsed laser excitation of EPL-375 pulsed laser.

**Figure S4. (a)**. Time resolved photoluminescence (TRPL) of dispersed $CsPbBr_3$ perovskite nanocrystals, showing radiative lifetime of 3.7ns. (b) Radiative lifetime of dispersed ternary halide perovskite nanocrystals, measured by TRPL.

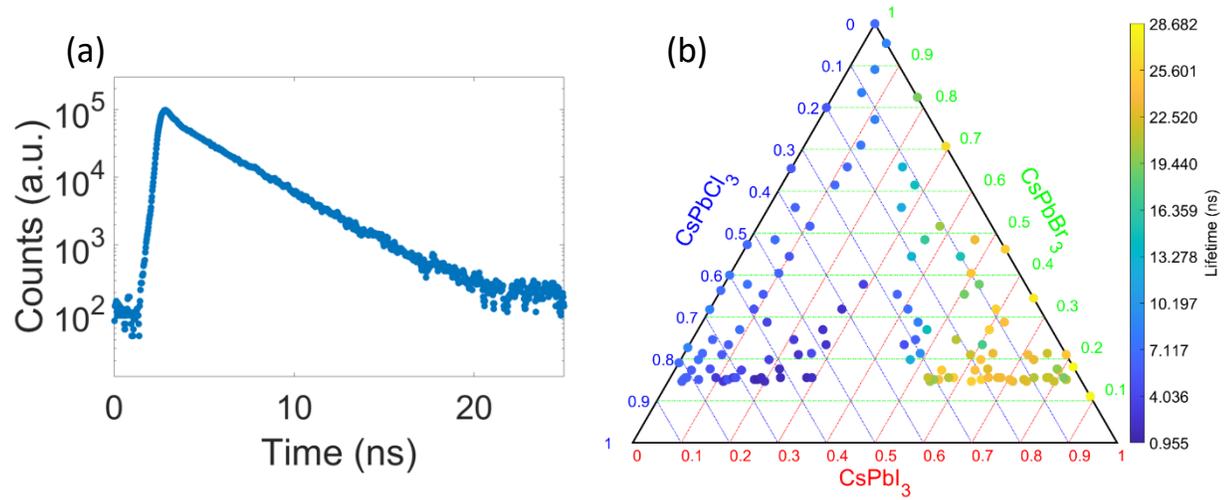

**Halide Composition Determination:**

In order to determine the halide composition of our nanocrystals we cross referenced optical calibration with high-throughput SEM-EDS measurements of selected samples. In the 12.4nm CsPbBr$_3$ NC sample, we determined the effective halide composition after the anion exchange by an optical calibration curve. Smaller nanocrystals are more complex for this method due to quantum confinement which is also dependent on the halide composition. In the binary Cl:Br and Br:I compositions, the PL of the alloy is linearly dependent on the halide compositions. The percentage of halide exchange in the X-site of the crystal (x), is calculated by:

$$x = \frac{|PL_x - PL_f|}{|PL_f - PL_i|} * 100\%$$

For the PL peak of the sample (PL$_x$), initial PL peak before the exchange (PL$_i$), and the PL of fully exchanged perovskite, from measurements of the pure single halide composition (PL$_f$).

Therefore, we used the PL of different samples with varied loadings of Cl or I to the halide composition at the X-site of the perovskite, as shown in Figure S5. These calibration binary halide calibration curves were extrapolated and used to determine the halide composition in the ternary halide systems. The ratio of Cl/I in the NC was determined by their concentration ratio in the exchange medium.

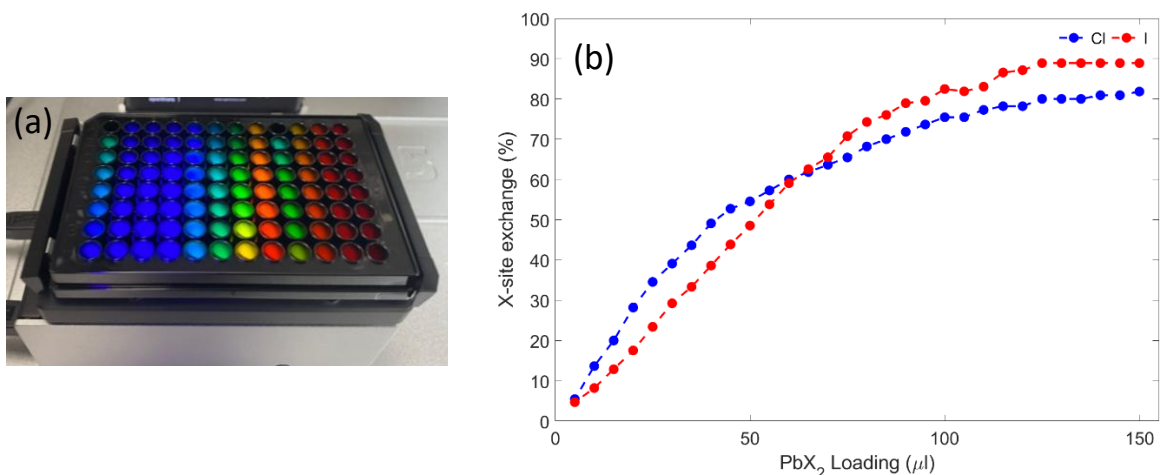

**Figure S5. (a)**. Image of high-throughput 96-well microplate loaded with binary and ternary halide perovskite nanocrystal dispersed samples under UV illumination. **(b)** X-site halide composition, as calculated by PL calibration curve of binary halide perovskite nanocrystals using 12nm CsPbBr$_3$ nanocrystals with different loadings of either PbCl$_2$ or PbI$_2$.

To confirm the optical calibration, we conducted scanning electron microscopy energy dispersive x-ray spectroscopy (SEM-EDS). EDS measurements for anion exchanged nanocrystals were carried out using Thermo Scientific Phenom XL G2 Desktop scanning electron microscope (SEM), equipped with 36 pin stub specimen mounts holder (shown in figure S6a). Samples of the deposited NCs films on Si wafer substrates were placed at a 10 mm working distance and were measured at an acceleration voltage of 15KV using the auto focus option. The EDS spectrum was measured using PhenomWorld elemental analysis software package from 5-15 points in each sample, and fitted (using ZAF correction) to contributions of the following elements: Cs, Pb, Cl, Br and I. First, we measured a set of 20 binary halide perovskite nanocrystals in SEM-EDS, as shown in Figure S6b-c. These results showed a median relative error of 10-15% from the PL calibration curves.

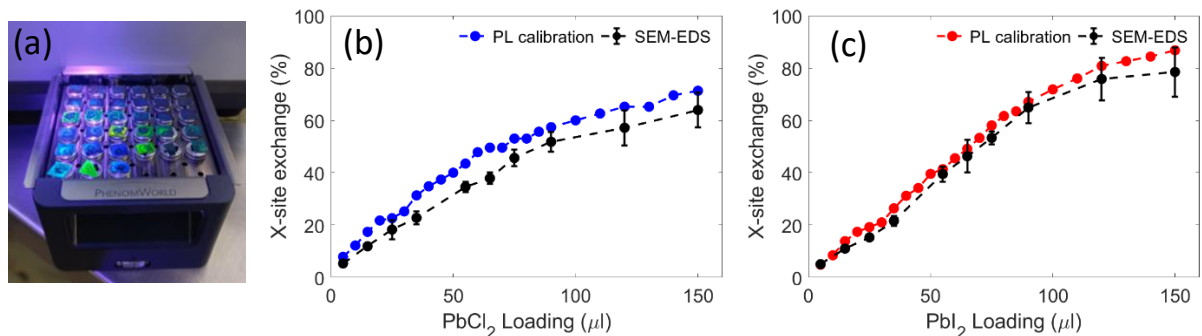

**Figure S6. (a)**. Image of high-throughput SEM sample holder loaded with binary and ternary halide perovskite nanocrystal samples on Si substrates under UV illumination. **(b-c)** X-site halide composition, as measured by SEM-EDS, of binary halide perovskite nanocrystals using 12nm $CsPbBr_3$ nanocrystals with different loadings of either (b) $PbCl_2$ or (c) $PbI_2$. The measured composition is compared to the expected composition by PL calibration curve.

EDS measurements of 38 samples of ternary halide perovskite nanocrystals are shown in Figure S7 for different total exchange solution loadings. As the optical calibration neglects any co-dependency of the exchange of one halide on the complimentary halide, measured compositions in higher total exchanges showed larger deviation from the optical calibration. The total median relative error was 14.8% from the optical calibration.

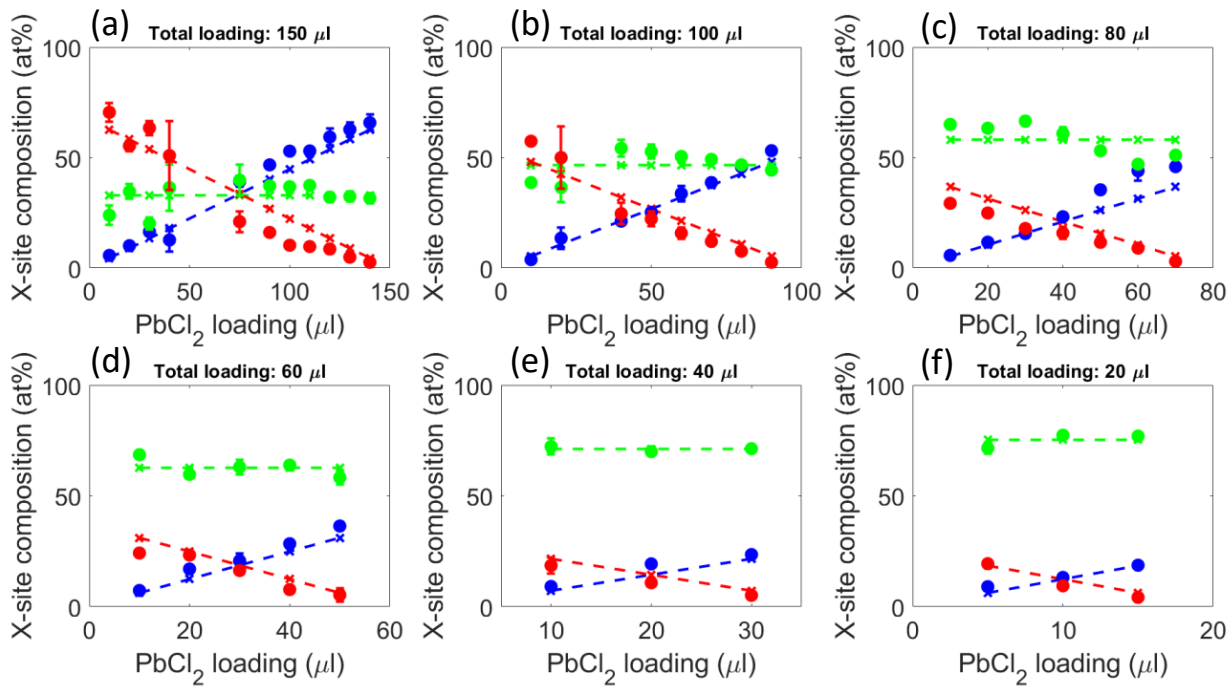

**Figure S7. (a-f)** X-site halide composition, as measured by SEM-EDS, of ternary halide perovskite nanocrystals using 12nm CsPbBr$_3$ nanocrystals with different loadings of both PbCl$_2$ and PbI$_2$ for total loading volumes of (a) 150µl, (b) 100µl, (c) 80µl, (d) 60µl, (e) 40µl, and (f) 20µl. The measured composition (colored points) of Cl (blue), Br (green) and I (red) is compared to the expected composition by extrapolated PL calibration curve (dashed lines).

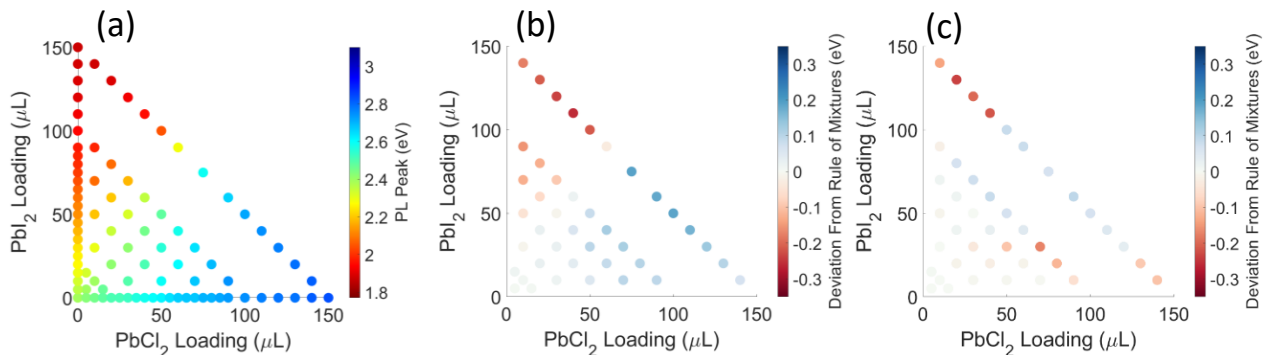

**Figure S8. (a)**. Emission maximum of binary and ternary halide perovskite nanocrystals, also measured by SEM-EDS. **(b)** Deviation from ideal solid-solution rule of mixture, using the PL peak presented in panel (a) by PL calibration curve. **(c)** Deviation from ideal solid-solution rule of mixture, using the PL peak presented in panel (a) by halide composition measured by EDS.

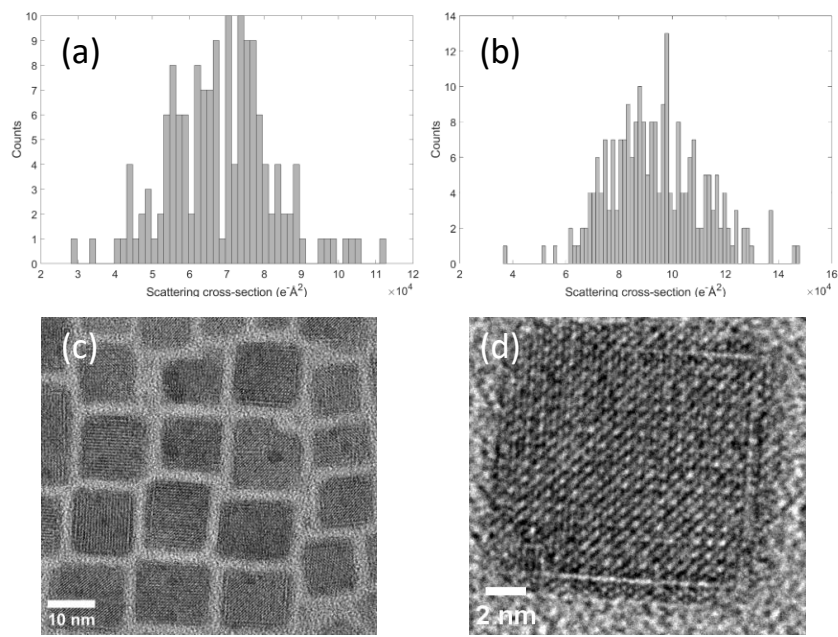

**Figure S9. (a-b)** histograms of calculated electron scattering cross section using StatSTEM software on atomic resolution HAADF-STEM micrographs presented in (a) Figure 4b-c and (b) Figure 4d-e. **(c-d)**. Additional TEM micrographs of ternary halide perovskite nanocrystals showing planar stacking fault defects.

## DFT calculations:

Using the Vienna Ab Initio Simulation Package (VASP)(*3*, *4*), first-principle structural relaxations are performed on bulk models (used to train a Cluster Expansion) and surface models. The Atomic Simulation Environment (ASE)(*5*) software package is used to interact with crystal structures and to intitiate the VASP calculator. The PerQueue (*6*)workflow manager is used to orchestrate the batch submission of relaxation jobs. Interactions between electrons and ions are described through Projector Augmented Waves (PAW)(*4*) pseudopotentials with an energy cut-off of 520 eV. Exchange-correlation effects are described through the Generalized Gradient Approximation (GGA) framework, using the PBEsol functional(*7*). All structural relaxations have been performed using a Γ-centered k-point grid with a grid density of ($4*2\pi$) k-points/Å. In surface models, only one k-point is sampled in the z-direction (the lattice parameter perpendicular to the surface). Partial orbital occupancy is treated with Gaussian smearing width of 0.05 eV. For the bulk models, all lattice parameters and ionic positions are allowed to relax. For the surface models, however, the lattice parameters and the atoms in the last single perovskite unitcell layer representing the bulk are fixed. In all cases, the conjugate gradient algorithm is used as implemented in VASP, until the total energy was converged to $10^{-6}$ eV and the residual forces on all atoms were below 0.05 eV/Å. All the input and output files generated in VASP associated with this project are stored in a repository and freely accessible on DTU Data.

## Cluster Expansion Model:

We use the CLEASE(*8*) software functionalities to build a Cluster Expansion model. The PerQueue workflow manager is used to autonomously orchestrate relaxation jobs and ground-state search and an evaluation of the ECI's, until the Cross Validation (CV) score threshold of 4 meV/atom is reached. We use polynomial basis functions and a maximum cluster diameter of 5 Å and a maximum cluster size of 4 atoms. In our CE model, the relaxed structural energy is determined and mapped on the initial lattice model.

To obtain a good resolution in the ternary phase diagram, a larger unit cell is needed than DFT calculations can be performed on. Therefore, we train the ECI values on systems of varying sizes to include some possible large-size effects that cannot be fully captured in a 40 atoms unit cell. The total dataset contains 120 structures of 40 atoms, 100 structures of 80 atoms, 175 structures of 160 atoms, 50 structures of 240 atoms and 30 structures of 360 atoms – a total of 475 data points which are all used to obtain the ECI values. Using the Leave-One-Out Cross Validation (LOOCV) method with l2 scoring scheme, the CV score reaches 3.860 meV/atom on this dataset. The root mean square error is 3.144 meV/atom, as can be seen in the model performance graph in Figure S10a. The contributions to the ECI values for different cluster sizes are shown in Figure S10c. The 0-body cluster (i.e. the background energy) contributes the most to the total energy. 1-body clusters (i.e. the species of a single atom) contribute to the total energy in the order of 0.1-0.3 eV/atom.

To find the ground state structure, a Monte Carlo (MC) annealing step is performed on 462 structures, using the found ECI values to assess the structural energy. Temperature is decreased from 10000000000 K to 10000, 6000, 4000, 2000, 1000, 800, 600, 400, 300, 200, 100, 50, 20 and 2 K using 1440000 MC sweeps per temperature.

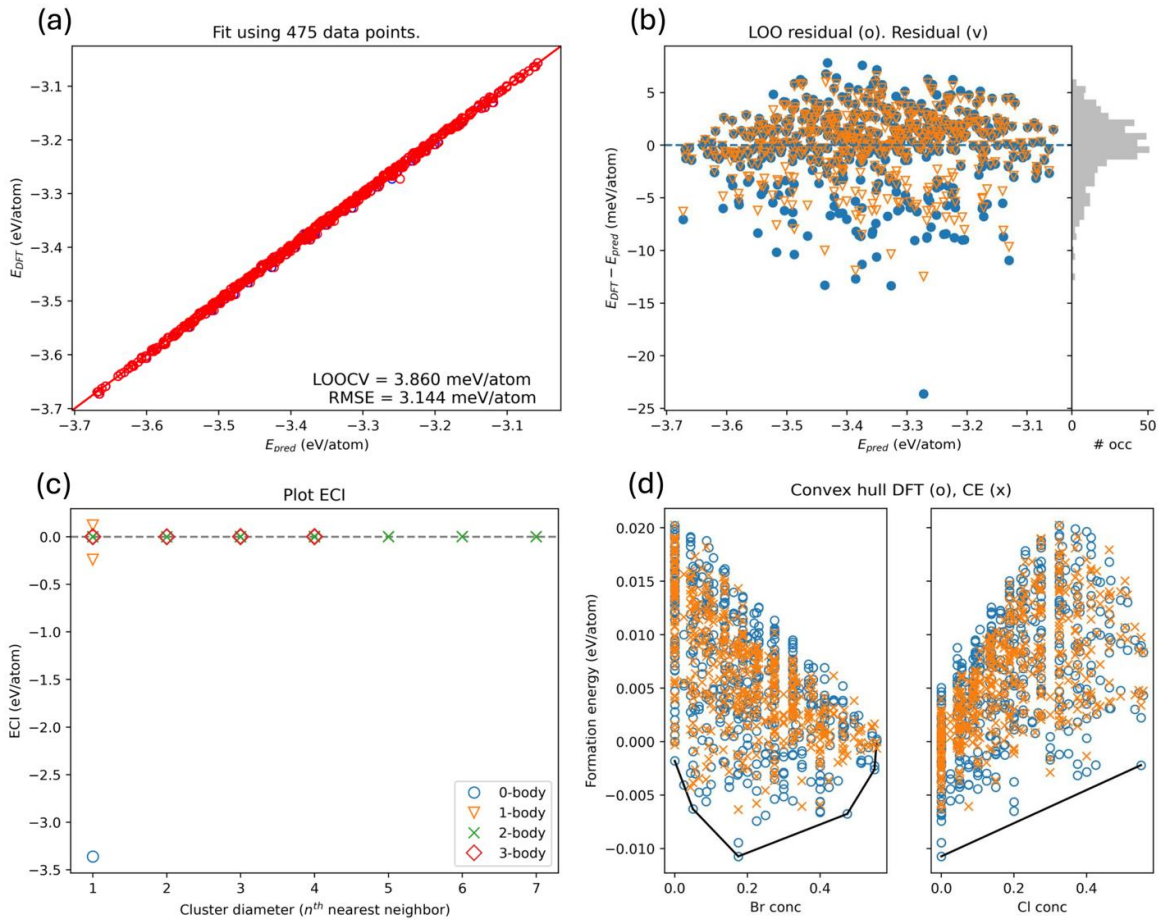

**Figure S10. Cluster expansion specifications on the training procedure. (a)** Fit of the DFT energy per structure in eV/atom vs the predicted energy by the ECI values, **(b)** residual plot, **(c)** The ECI value per cluster diameter for different cluster sizes, **(d)** Convex hull as a function of the Br and Cl concentration. Both DFT values and CE predictions are plotted.

## Surfaces DFT Model

Surface models are generated with the ASE software. We use the rule of mixtures to obtain lattice parameters for the surface slab based on the percentage of different halides present. First, we investigate if a small concentration (25%) of I in a $CsPbCl_3$ host lattice will show the lowest energy when I clusters on the surface, or in the middle of the slab. A visual representation of these test models is shown in Figure S11a. We find a total energy of -3.496 eV/atom when Iodine clusters on the surface of the slab, compared to -3.464 eV/atom when Iodine orients in the middle of the slab. Therefore, we proceed with halide-species substitution at the surface of a model slab for the rest of our calculations.

A gradient of the guest species in the layers is introduced in a stepwise manner into the base lattice, under a homogeneous background of Br introduced in layer 1-6. In Figure S11b and S11c, we show how the structural characteristics for each datapoint in $CsPbCl_{1.875}Br_{0.75}I_{0.375}$ and $CsPbCl_{0.75}Br_{0.75}I_{1.50}$ respectively.

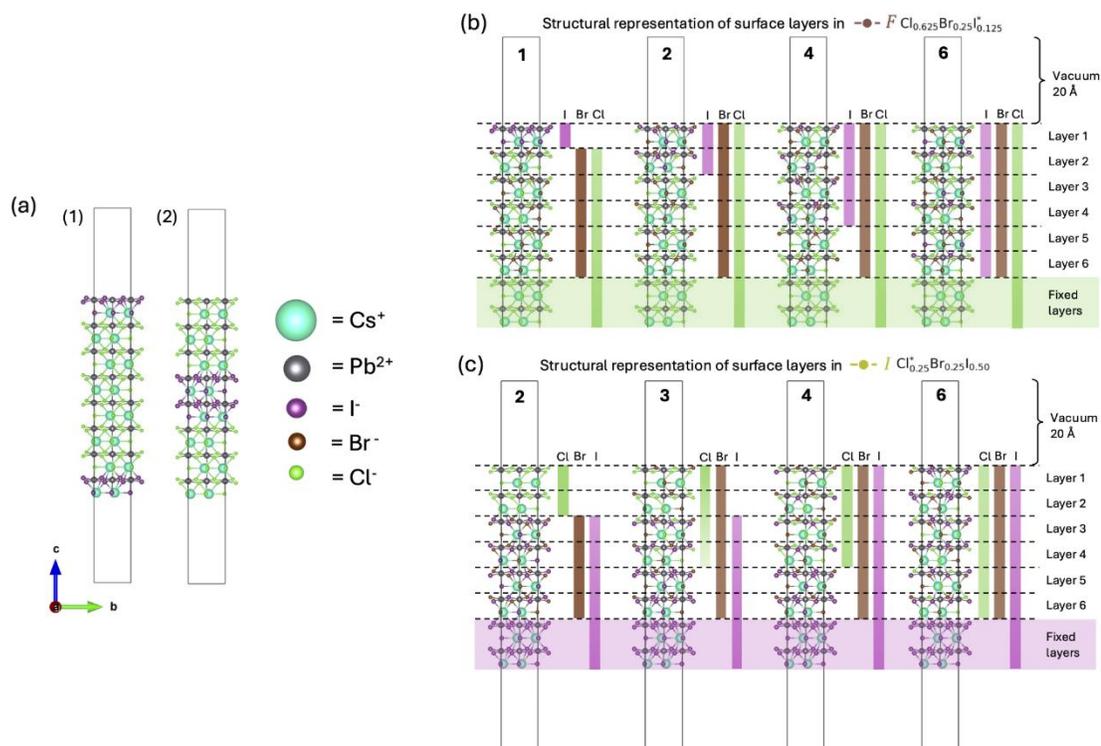

**Figure S11. Surface models.** (a) Surface models to perform the test. In a-(1), a CsPbCl3 structure where 12.5% I is added only to the outside of the slab, and in a-(2), a CsPbCl3 structure where 12.5% I is added only to the inside of the slab. **(b-c)** Surface models of $CsPbCl_{1.875}Br_{0.75}I_{0.375}$ and $CsPbCl_{0.75}Br_{0.75}I_{1.5}$ respectively. Numbering of the layers is given on the side. The presence of different halide species in each layer is indicated with the coloured bars on the side of each structure, where the opacity is a measurement for the concentration across the layers. Number 1, 2, 4 and 6 (Figure b) and 2, 3, 4 and 6 (Figure c) indicate the penetration layer of the guest species and correspond to the x-axis in Figure 4c-e in the main text.

## Robotic System Precision and Reproducibility

The random error of pipetting in the OT-2 robotic system spans from ±0.6% for 300 µl to ±15% for 1 µl (minimal operable volume) according to the specifications. The estimated random error for the minimal volume used in this study of 5 µl was ±4%, and for 50 µl – around ±2%.

To check the reproducibility experimentally, 50 µl of $CsPbBr_3$ nanocrystals were dispensed into each well of a 96-well plate (except for two blank wells A1 and H12). Spectroscopic characteristics were measured (Figure S12). PL intensity total spread was within ±4% (Figure S12(a,b)), which corresponds well with pipetting specifications. Spread for the measured absorption was noticeably higher, ±9% (Figure S12(c,d)). The evident absence of a positive correlation between the absorption and emission values (Figure S13) indicates that the error is not due to pipetting error. Therefore, the apparent reason for the error in the absorption measurement is the measurement error, caused by the fact that the NCs solution used for this experiment was diluted 10 times compared to the solution used in the other experiments. Thus, the presented error is an upper limit, and the actual irreproducibility in the real experiment was lower.

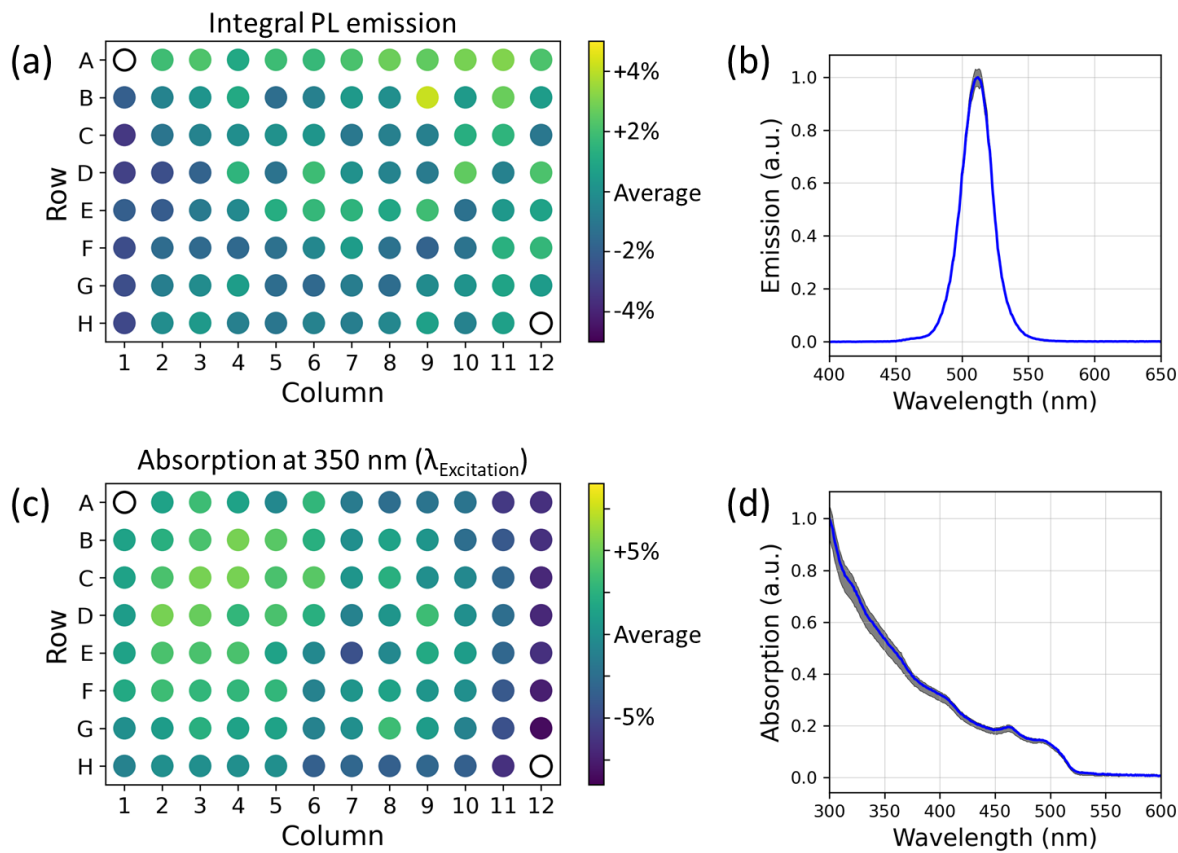

**Figure S12. (a)** Peak values of the PL emission intensity for a 96-well plate, loaded with a similar amount of $CsPbBr_3$ nanocrystals, and **(b)** emission spectra – the blue line corresponds to the averaged emission, and the grey shaded area shows the spread from the lowest to the highest values. **(c)** Values of absorption at 350 nm and **(d)** absorption spectra for the same well.

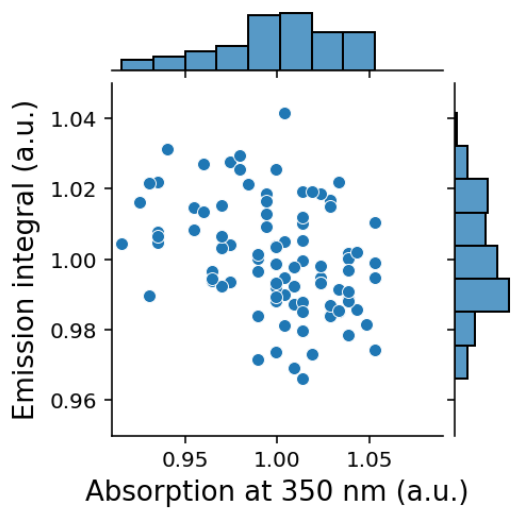

**Figure S13.** Scatter plot between optical density at 350 nm and integral PL emission intensity for a 96-well plate, loaded with a similar amount of $CsPbBr_3$ nanocrystals.